\documentclass{IOS-Book-Article}

\usepackage{mathptmx}
\usepackage{forest}
\usepackage{adjustbox}
\usepackage{soul}\setuldepth{article}
\usepackage{enumitem}

\def\hb{\hbox to 11.5 cm{}}

\begin{document}

\pagestyle{headings}
\def\thepage{}
\begin{frontmatter}              


\title{
The Data Sharing Paradox of Synthetic Data in Healthcare
}

\markboth{}{\today \hb}
\author[A]{\fnms{Jim} \snm{Achterberg}\orcid{https://orcid.org/0009-0000-
9589-7831} \thanks{Corresponding Author: Jim Achterberg, j.l.achterberg@lumc.nl}},
\author[A]{\fnms{Bram} \snm{van Dijk}\orcid{https://orcid.org/0009-0002-9176-1608}}
\author[B]{\fnms{Saif ul} \snm{Islam}\orcid{https://orcid.org/0000-0002-
9546-4195}}
\author[B]{\fnms{Hafiz} \snm{Muhammad Waseem}\orcid{https://orcid.org/0000-0002-9418-1492}},
\author[C]{\fnms{Parisis} \snm{Gallos}\orcid{https://orcid.org/0000-0002-8630-7200}},
\author[B]{\fnms{Gregory} \snm{Epiphaniou}\orcid{https://orcid.org/0000-0003-1054-6368}},
\author[B]{\fnms{Carsten} \snm{Maple}\orcid{https://orcid.org/0000-0002-4715-212X}},
\author[A]{\fnms{Marcel} \snm{Haas}\orcid{https://orcid.org/0000-0003-
2581-8370}}
and 
\author[A,D]{\fnms{Marco} \snm{Spruit}\orcid{https://orcid.org/0000-0002-9237-221X}}

\address[A]{Leiden University Medical Center, The Netherlands}
\address[B]{University of Warwick, UK}
\address[C]{European Federation for Medical Informatics, Switzerland}
\address[D]{Leiden Institute of Advanced Computer Science, Leiden University, The Netherlands}

\begin{abstract}
Synthetic data offers a promising solution to privacy concerns in healthcare by generating useful datasets in a privacy-aware manner. However, although synthetic data is typically developed with the intention of sharing said data, ambiguous reidentification risk assessments often prevent synthetic data from seeing the light of day. One of the main causes is that privacy metrics for synthetic data, which inform on reidentification risks, are not well-aligned with practical requirements and regulations regarding data sharing in healthcare. This article discusses the paradoxical situation where synthetic data is designed for data sharing but is often still restricted. We also discuss how the field should move forward to mitigate this issue. 

\end{abstract}

\end{frontmatter}
\markboth{\today \hb}{\today \hb}

\section{Introduction}

Healthcare data, characterized by its multidimensional nature and diverse formats—such as images, text, tables, time series, and videos poses significant challenges for ensuring privacy. Synthetic Data (SD) aims to replicate the complex relationships among features within such data while mitigating the risk of reidentification. However, most existing privacy metrics struggle to navigate this complexity. 
The practical requirements on patient privacy, stemming from relevant regulatory frameworks, are typically poorly captured by the metrics used to assess SD privacy. Because of this, it is often still unclear whether SD satisfies these requirements and publishing SD is often forgone. Here we discuss this mismatch between practical requirements and quantitative metrics for privacy, and mention several avenues of exploration that could help to reconcile the two.

\section{Regulatory requirements for data anonymity}

Whether health data can be published depends upon relevant regulatory frameworks like the GDPR in the EU and HIPAA in the US. Both state that data can be published if it is \textbf{anonymous}, meaning there is no risk of reidentification:
\begin{itemize}
    \item GDPR, Recital 26: ``The principles of data protection should therefore not apply to anonymous information, namely information which does not relate to an identified or identifiable natural person or to personal data rendered anonymous in such a manner that the data subject is not or no longer identifiable."
    \item HIPAA, 45 CFR § 164.514 ``Health information that does not identify an individual and with respect to which there is no reasonable basis to believe that the information can be used to identify an individual is not individually identifiable health information."
\end{itemize}

\noindent Although SD is typically generated without a one-to-one mapping to Real Data (RD), SD is not inherently anonymous  \cite{jordon2022synthetic,stadler2022synthetic}; only in SD sufficiently different from the RD it was based on, privacy guarantees are strong. Simultaneously, dissimilarity between RD and SD undermines its very use, hence privacy and utility must be traded off. This makes SD in healthcare subject to the same strict criteria as RD with regard to data sharing. 

To minimise reidentification risk, SD generating parties resort to privacy-preserving techniques when generating SD. Here, reidentification risk means inference of an individual's sensitive attributes or membership to the training set of RD used to generate SD \cite{osorio2024privacy}. Yet, it is currently unclear whether minimizing reidentification risk can render SD anonymous, whether SD can be better understood as pseudonymized data, and what other measures must be taken to prevent reidentification through SD in the future \cite{fontanillo2022synthetic,lopez2022legal}. 

In addition, such questions are virtually impossible to address without context. In the case of a privacy attack, an \textit{attacker} typically tries to disclose (sensitive) information from the SD or the SD generating model about some other \textit{actor} \cite{rigaki2023survey}. An attacker could be a health insurer which can combine data from their clients with published SD on some clinical trial, and infer which of their clients took part in the trial since they are likely contained in the SD training set. Since contexts vary, privacy risk is not an absolute measure but changes as events, times, and actors change \cite{lopez2022legal}. Hence, the question of whether SD provides anonymous, pseudonymous, or personal data depends on many factors. For example, it has been argued that reidentification risk could render SD generating algorithms personal data, so these algorithms must be GDPR and HIPAA compliant as well \cite{veale2018algorithms}, underscoring that the issue of privacy in SD is far from resolved.    

\section{Privacy metrics in synthetic data}
Since re-identifiable data cannot be published under relevant regulations (e.g., GDPR or HIPAA), SD intended for sharing must be evaluated for its reidentification risk. Two main categories of privacy metrics can be distinguished, assuming an attacker has some information about an individual: \textbf{attribute disclosure} and \textbf{membership disclosure} \cite{jordon2022synthetic}.

Attribute disclosure metrics indicate the risk of an attacker inferring real, identifiable information from SD. This requires some subset of real variables to be available to the actors to link from synthetic to real information. Attackers can then link synthetic to real information through e.g., prediction by training an inference model on SD \cite{stadler2022synthetic}, or through similarity in terms of statistical distance \cite{kim2024synthesis}. Here, a wide variety of prediction models or statistical distance measures might be employed. The inference accuracy of the real, identifiable information is an indication of reidentification risk. 

Membership disclosure metrics indicate the risk of attackers inferring which individuals were contained in the training set of an SD generator \cite{shokri2017membership}. This is especially relevant when membership to the RD exposes identifiable information and thereby invokes reidentification risk. For example, for SD generated from a hospital's pulmonary department, an attacker may infer if an individual was a patient at said department. Usually, such inference requires the attacker to have access to a set of real personal data. This might occur when an individual has their own data agreement with the attacker, for example, between a patient and a health insurance company. In this case, the attacker has access to a large amount of identifiable information and may try to use this to infer even more sensitive information. To perform membership inference, the attacker estimates the likelihood that an individual comes from the SD distribution. If this is relatively likely, e.g., compared to the distribution of the real information he has access to, the individual is likely to have been contained in the training set \cite{van2023membership}. In scenarios where membership to the synthetic dataset is deemed sensitive, the membership inference accuracy indicates the risk of re-identifying said sensitive attribute.

\section{Challenges in privacy assessments of synthetic data}
Both attribute and membership disclosure metrics can inform on the reidentification risk from SD. However, several issues persist, causing uncertainty around whether SD is truly anonymous and is allowed to be shared.

Firstly, there are multiple ways of assessing privacy from a legal perspective. The `zero-risk' approach in privacy assessment requires an SD generating party to take all possible attackers and methods that can be employed to attempt reidentification into account in the present and future. This can be challenging, given that it is often unknown which technologies or other datasets attackers can access. A more pragmatic approach, also known as the `acceptable risk' approach, estimates reidentification risk by focusing only on likely attackers and methods that are, for example, similar and known to the SD generating party \cite{lopez2022legal}. Still, the question of which approach is most appropriate in which context is open; civil law tends to defend zero-risk approaches (with the GDPR as the most common example), and common law tends to defend acceptable-risk approaches, which may confuse SD generating parties and prevent them from publishing their SD \cite{lopez2022legal}. Moreover, legal precedents exist mainly for older privacy metrics like k-anonymity \cite{osorio2024privacy}, but not for more recent frameworks like differential privacy, so SD generating parties will often feel forced to stick with the old.        

Secondly, although the risk of reidentification can be estimated through, e.g., disclosure of attributes and membership to training data, there are no shared thresholds in the SD-generating community to indicate when this risk is \textit{non-existent}. This has great implications since the reidentification risk must be zero to render SD truly anonymous and thus shareable.

Third, many privacy metrics were not originally designed to address healthcare-specific use cases. Techniques such as k-anonymity and l-diversity focus primarily on preventing direct reidentification \cite{osorio2024privacy} but may fall short in healthcare contexts where data complexity and sensitivity are elevated. Differential privacy, while theoretically robust, can significantly alter the utility of data. In addition, it may not protect against all types of privacy attacks, particularly in complex datasets such as electronic health records.

Fourth, ethical considerations further complicate the use of SD in healthcare. Patients expect control over how their data is used, even when anonymized or synthesized. Although privacy metrics address technical risks, they often overlook ethical issues surrounding data use in research, clinical trials, or healthcare applications and are hard to explain in lay terms. Differential privacy, for example, can introduce biases that disproportionately affect under-represented groups, leading to skewed clinical decisions or discriminatory outcomes \cite{dankar2013practicing}.


Lastly, the technical literature on the assessment of SD privacy has been largely oblivious to the evolving context of SD. Attackers' knowledge about individuals can increase over time, and they may be able to exploit more public datasets as these become increasingly available, which impacts reidentification risk. In addition, regulating bodies can place more precise restrictions on SD to protect sensitive information for particular time periods, but current privacy metrics do not take this into account.

\section{Recommendations}
Several approaches are being explored to bridge the gap between privacy requirements and metrics in synthetic healthcare data. These include:
\begin{itemize}
    \item \textbf{Context-aware privacy metrics:} Privacy requirements are often context-specific, as information may be more or less sensitive depending on factors such as the relevant domain, availability of external knowledge, and the potential gain to attackers. To better reconcile privacy requirements and metrics in different scenarios, it may thus be helpful to focus efforts towards developing context-specific rather than general-purpose privacy metrics.
    
    \item \textbf{Explainable privacy assessments:} Data privacy is increasingly important for both patients and healthcare providers, and they may show more privacy-protective behaviour if the protection of their information depends on arcane metrics that are hard to explain. That is, current privacy metrics must be explained in more familiar notions that resonate with the broader public \cite{dankar2013practicing}.
    \item \textbf{Exchanging knowledge:} Academic researchers have to get in touch with, e.g., policymakers and legal experts, to better explain current technical privacy-preserving approaches so that legal precedents and law can be created that enable the adoption of privacy metrics in practice. In addition, policymakers and legal experts would benefit from expertise from the SD generating parties to get a clearer view of what `acceptable risks' are, which depends on ongoing research in models generating SD, privacy metrics, and other technology.
\end{itemize}

\section{Conclusion and Discussion}
The mismatch between practical requirements on privacy for data sharing and relevant quantitative risk metrics are major causes of the stagnating adoption of synthetic data in healthcare. To address this issue, the scientific community must take action by developing more suitable privacy metrics, enhancing the understanding of those metrics to the general public, and exchanging knowledge with policymakers and legal experts so that relevant thresholds and legal precedents can be developed.

\section{Acknowledgements}\label{sec:acknowledgement}
This work is co-funded by the HORIZON.2.1 - Health Programme of the European Commission, Grant Agreement number: 101095661 - Innovative applications of assessment and assurance of data and synthetic data for regulatory decision support (INSAFEDARE).

\bibliographystyle{vancouver} 

\end{document}